\documentclass[conference]{IEEEtran}
\IEEEoverridecommandlockouts
\usepackage{cite}
\usepackage{amsmath,amssymb,amsfonts}
\usepackage{algorithmic}
\usepackage{graphicx}
\usepackage{textcomp}
\usepackage{xcolor}
\usepackage{multirow}
\usepackage{listings}
\usepackage{adjustbox}
\usepackage{float}
\usepackage{pifont}
\usepackage{balance}
\usepackage{todonotes}

\def\BibTeX{{\rm B\kern-.05em{\sc i\kern-.025em b}\kern-.08em
    T\kern-.1667em\lower.7ex\hbox{E}\kern-.125emX}}

\def\code#1{\texttt{#1}}
\begin{document}

\title{Devils in the Clouds: An Evolutionary Study of Telnet Bot Loaders
% \thanks{Identify applicable funding agency here. If none, delete this.}
}
% \author{Anonymous authors}

\author{\IEEEauthorblockN{
Yuhui Zhu\IEEEauthorrefmark{1}\IEEEauthorrefmark{2},
Zhenxiang Chen \IEEEauthorrefmark{1}\IEEEauthorrefmark{2}\IEEEauthorrefmark{6},
Qiben Yan\IEEEauthorrefmark{3},
Shanshan Wang\IEEEauthorrefmark{1}\IEEEauthorrefmark{2},
Alberto Giaretta\IEEEauthorrefmark{4},
}

\IEEEauthorblockN{
Enlong Li\IEEEauthorrefmark{1}\IEEEauthorrefmark{2},
Lizhi Peng \IEEEauthorrefmark{1}\IEEEauthorrefmark{2},
Chuan Zhao \IEEEauthorrefmark{1}\IEEEauthorrefmark{2},
Mauro Conti\IEEEauthorrefmark{5}
}

\IEEEauthorblockA{\IEEEauthorrefmark{1}Shandong Provincial Key Laboratory of Network Based Intelligent Computing, University of Jinan, China}
\IEEEauthorblockA{\IEEEauthorrefmark{2}School of Information Science and Engineering, University of Jinan, China}
\IEEEauthorblockA{\IEEEauthorrefmark{3}Department of Computer Science and Engineering, Michigan State University, USA}
\IEEEauthorblockA{\IEEEauthorrefmark{4}AASS MPI Lab, Örebro University, Sweden}
\IEEEauthorblockA{\IEEEauthorrefmark{5}Department of Mathematics, University of Padua, Italy}
\IEEEauthorblockA{\IEEEauthorrefmark{6}Corresponding author, Email: czx.ujn@gmail.com}
}

\maketitle

\begin{abstract}
    % !TeX root = ..\main.tex
One of the innovations brought by Mirai and its derived malware is the adoption of self-contained loaders for infecting IoT devices and recruiting them in botnets.
Functionally decoupled from other botnet components and not embedded in the payload, loaders cannot be analysed using conventional approaches that rely on honeypots for capturing samples.
Different approaches are necessary for studying the loaders evolution and defining a genealogy.
To address the insufficient knowledge about loaders' lineage in existing studies, in this paper, we propose a semantic-aware method to measure, categorize, and compare different loader servers, with the goal of highlighting their evolution, independent from the payload evolution.
Leveraging behavior-based metrics, we cluster the discovered loaders and define eight families to determine the genealogy and draw a homology map.
Our study shows that the source code of Mirai is evolving and spawning new botnets with new capabilities, both on the client side and the server side. In turn, shedding light on the infection loaders can help the cybersecurity community to improve detection and prevention tools.

%Self-contained loaders have been widely adopted by Mirai and its descendants for exploiting vulnerabilities and spawning new bots. 
%As they are seldomly distributed via remote infections and functionally decoupled with other botnet components, conventional sample-based perspectives and methodologies are not applicable to depict the genealogy of these infrastructure components.
% =======================================
% To complement the biaevolving with novel additions on both the client and server side, thereby offering a new perspective on finding covert relationships among cybercrimes based on the botnet infrastructures.sed knowledge of bot loaders, we propose a semantic-aware method to measure, categorize, and compare the functions of loader servers.
% Leveraging behavior-based metrics, we cluster the observed loaders and define eight families to determine the genealogy and homology of their infection toolkits.
% While the client-side threat intelligence is disjoint from server-side dynamics, our work reveals the evolving trend of server-side bot loaders. 
% %which is similar to bot clients.
% This further corroborates that the released code of Mirai keeps spawning new generations of botnets with novel additions on both the client and server side, thereby offering a new perspective on finding covert relationships among cybercrimes based on the botnet infrastructures.
 
\end{abstract}


\section{Introduction}
% !TeX root = ..\main.tex
Following the growth of the IoT market, botnets recruiting IoT devices have become a major cyber-security threat.
Earlier in 2016, the emerging Mirai botnet drew attention from the cybersecurity community.
The operator launched a 1.1Tbps DDoS attack using 148,000 IoT devices, breaking the record and making it the most notorious botnet clan in the following years.
As the Mirai's code release~\cite{anna-senpaiFREEWorldLargest2016} have stimulated the evolution of botnet malware, the cybersecurity community has invested considerable energy to define complete taxonomies that would help to understand the differences and similarities between emerging variants.
Empirical studies have discovered multiple \textit{families} and defined by general taxonomy studies~\cite{antonakakisUnderstandingMiraiBotnet2017,herwigMeasurementAnalysisHajime2019,griffioenExaminingMiraiBattle2020}, while other efforts went into collecting and dissecting malware samples to identify less obvious \textit{variants}~\cite{cozziTangledGenealogyIoT2020,downingDeepReflectDiscoveringMalicious2021,wangEvolutionaryStudyIoT2021}.

Although these investigations depicted a clear genealogy of malware families and variants, sample-centered approaches fail to consider server-side components that are characteristics of Mirai-like botnets. In particular, while conventional worms infect new victims independently, Mirai exhibits a decoupled design that assigns infection functions to a separated self-contained \emph{loader server}, deployed on cloud services. 
By assigning to the bots the discovery task and offloading the infection process to an external loader server, Mirai reduces the amount of resources required for a machine to function as a bot and allows for recruiting resource-restricted IoT devices.
%less resource-consuming.

The design choice of Mirai presents different challenges for the research community. On the one hand, delegating the infection tasks to cloud services results in botnets that are split in disjoint parts, making them harder to study as a whole system. On the other hand, this choice prevents honeypots from capturing a vital part of Mirai code and operations.
Therefore, studies limited to payload-based lineage inference are inherently incomplete, as they neglect the infecting toolkits. Server-side studies are critical to understand in depth botnets infrastructure. 
By highlighting the peculiar characteristics of intrusion toolkits, malware studies can shed new light on the relationship between botnet campaigns, as well as improve the efficiency of defense strategies against new attack vectors.
%It also sheds light on botnet infrastructures and helps the cybersecurity community understand the relationships between botnet campaigns from a new perspective. 

%Given this asynchronous development, to analyse the evolution of botnets and infection malware it is insufficient to focus on the payloads and neglect the infecting infrastructure.
%discussing the variation and evolution of infection toolkits, the asynchronized development of bot clients and loaders makes it inappropriate to simply drive the knowledge of families from malware samples.
%Moreover, as these cloud-based loaders are less distributed than the bot clients, the sample-based lineage inference is also infeasible due to the limitation of honeypots.
%The variation of exploitable vulnerabilities not only catalyzed the evolution of loaders' intrusion toolkits but also demonstrated the urgency of server-side studies of botnets.

In this paper, we focus on the telnet loader, the only infection toolkit distributed in the original Mirai codebase.
Our work provides a server-side view of botnets evolution and a novel behavior-based taxonomy of bot loaders, following a conventional family-variant epistemology.
%with absent loader samples.
To address the absence of loader samples, we analyse the interaction logs captured using telnet honeypots. Under the assumption that different intrusion toolkits use different infection instructions,
%as objective representations of intrusion toolkits.
we adopt a semantic-aware strategy to map differences and similarities in instruction sets to lineages.
The paper's contributions are the following:
\begin{itemize}
  \item
    %While the server-side perspective is commonly overlooked due to the absence of samples, we
    We propose a semantic-aware method to analyse the lineage of bot loaders through their interaction logs, captured via honeypots;
  \item
    We conduct a taxonomy study on infection toolkits, evaluate the behavioural patterns, and define a genealogy of eight families;
  \item 
    We highlight the existence of an unconventional loader, suspected to conduct fileless attacks, and we confirm its homology with conventional file-based bot loaders;
  \item  
    We highlight that infection loaders evolve independently from their payloads and we advocate the importance of a server-side perspective in botnet provenance attribution.
\end{itemize}

\section{Related Work}
\label{sec_background}
% !TeX root = ..\main.tex
Since the source code of Mirai has been publicly released, dozens of variants appeared in the wild and hundreds of massive botnets spawned. The cybersecurity community strove to produce a taxonomic view on these botnet campaigns and capture their evolution.
%As the released code of Mirai and other botnets have spawned massive botnets, the cybersecurity community has attempted to offer a taxonomic view on massive botnet campaigns and systematically scrutinize their evolution.
Most studies built their observations on the relationships between botnets on empirical definitions of \emph{families} and \emph{variants}.
Pa et al. \cite{paIoTPOTAnalysingRise2015} analyzed and categorized emerging botnets based on an observation of shared command sequence patterns.
Antonakakis et al.~\cite{antonakakisUnderstandingMiraiBotnet2017} and Herwig et al.~\cite{herwigMeasurementAnalysisHajime2019} analyzed two emerging botnet families, Mirai and Hajime, to discuss their propagation and evolution.
To examine the competition and battle among botnets, Griffioen~\cite{griffioenExaminingMiraiBattle2020} categorized botnet campaigns into several variants by their identity strings before discussing their behavior.
Dang et al.~\cite{dangUnderstandingFilelessAttacks2019} categorized fileless attacks on Linux-based IoT devices and correlated these attack vectors with known botnet families.
Alrawi et al.~\cite{alrawiCircleLifeLargeScale2021} discussed the lifecycles of botnets based on family definitions from VirusTotal~\cite{VirusTotal}, a popular malware detection tool with a collection of anti-virus engines.

Beyond the definition of families, depicting their evolution and variation under the family-variant epistemology is also critical to botnet studies.
Most studies obtain evidence from bot samples, the most easy-to-access components, by involving \emph{bindiff} and other sample-centered techniques.
Wang et al.~\cite{wangEvolutionaryStudyIoT2021} pointed out that investigating relationships among botnet families could be a fundamental step for provenance, triage, labeling, lineage analysis, and authorship attribution.
They derived knowledge of botnet samples from online articles and captured samples, proposed a hybrid methodology to construct a lineage graph, and discussed the lineage of 72 botnet families.
Cozzi et al.~\cite{cozziTangledGenealogyIoT2020} shed light on the tangled genealogy of botnet samples by measuring shared components across malware samples from different families.
Most anti-virus engines also used YARA~\cite{YARA} to match the shared patterns of a malware family, so that they could relate unseen samples to existing families or variants.

As dissecting samples of bot loaders is impractical due to the absence of samples, many studies further explored various intrusion fingerprints to reveal their covert relationships.
The first studies on Mirai~\cite{antonakakisUnderstandingMiraiBotnet2017}, and Hajime~\cite{herwigMeasurementAnalysisHajime2019} studied the password dictionaries captured by honeypots to discuss the lineage of botnet variants.
Lingenfelter et al.~\cite{lingenfelterAnalyzingVariationIoT2020} made a comparison of initial commands and query tokens to demonstrate the variation of telnet intrusion toolkits.
Torabi et al.~\cite{torabiStringsBasedSimilarityAnalysis2021} tried mining unique strings from logs to build associations among active botnets.
Tabari et al.~\cite{tabariWhatAreAttackers2021} made a statistical analysis of the most commonly exploited vulnerabilities, credentials, and intrusion commands.
However, while sample-centered works have always overlooked loaders in lineage inference, simply comparing strings or fingerprints in intrusion toolkits cannot yield a systematic view.
The families and variants derived from malware samples also brought a prior hypothesis bias to these studies, which effectively hinders the understanding of  the desired server-side behaviors.

\section{Methodology}
\label{sec_methodology}
% !TeX root = ..\main.tex

% \begin{figure*}[htb]
%   \centering
%   \includegraphics[width=0.9\textwidth]{figures/methodology.pdf}
%   \caption{Schematic depiction of the analysis steps in our work.}
%   \label{fig_methodology}
%   % \Description[<short description>]{<long description>}
% \end{figure*}

In this section, we categorize telnet loaders into families and discuss their variation through captured sequences of intrusion commands.
% Why categorize??
We assume that the sequence of intrusion commands may reflect loaders' functions and inner implementation, so we propose a semantic-aware metric to describe the similarity among collected sequences and leverage agglomerative clustering to categorize them into families.
Based on the agglomerative tree, we further present the shared patterns among sibling loaders to yield a systematic conclusion about the variation and homology of intrusion toolkits from a server-side perspective.
% The complete process is depicted in Fig. \ref{fig_methodology}.

\subsection{Data Collection}

Telnet is a text-based protocol commonly used for accessing a remote shell on IoT platforms.
Although botnets have been evolving their toolkits to exploit new vulnerabilities in different protocols, bot masters are still working on telnet-based intrusions to adapt to more vulnerable devices.
Based on such behaviors, recent studies on IoT botnets \cite{griffioenExaminingMiraiBattle2020,herwigMeasurementAnalysisHajime2019,torabiStringsBasedSimilarityAnalysis2021} all considered telnet loaders as a crucial basis to make comparisons among botnet families.
Thus, we initialize the study by investigating telnet protocols to understand the behavior of botnet loaders.

We deploy a honeycloud system to record command sequences from loaders.
We deploy frontends on 3 virtiual cloud servers in China, Singapore, and the United States to redirect requests to the honeycloud backend.
The honeycloud backend dispatches telnet conversations to the backing devices listed in Table \ref{tab_device}, then it substitutes the requested username and password to allow botnets to access our deployed devices.
We only record the requests of intrusion commands from botnets and drop all responses to avoid client-side noise for our server-side analysis.
All requests collected from a conversation are concatenated into a single ``request log" to represent the behavior and function of a loader.

% !TeX root = ..\main.tex
% Please add the following required packages to your document preamble:
% \usepackage{multirow}
\begin{table}[h]
    \caption{Deployed backing devices during the experiment}
    \centering
    \begin{tabular}{ccc}
    \hline
    \textbf{Type}                          & \textbf{Device name} & \textbf{Software version}     \\ \hline
    \multirow{3}{*}{\textbf{Smart router}} & Lenovo Y1S           & PandoraBox git-6fcbaa5        \\ %\cline{2-3} 
                                           & Netgear R7800        & OpenWRT 21.02                 \\ %\cline{2-3}
                                           & Netgear R6300v2      & KoolShare Merlin              \\ \hline
    \textbf{IP Camera}                     & Hikvision            & (Stock)                       \\ \hline
    \textbf{ONU}                           & CMCC I-120EM         & (Stock)                       \\ \hline
    \textbf{Other}                         & Raspberry Pi 3B      & Raspberry Pi OS Lite Jan 2021 \\ 
    \hline
    \end{tabular}
    \label{tab_device}
\end{table}

\label{subsec_ctrl_group}
To evaluate the effectiveness of our proposed method, we run a Hajime bot and a Mirai loader in a QEMU ARM sandbox to generate control group data.
The Hajime bot sample is provided by MalwareBazaar\footnote{https://bazaar.abuse.ch/}.

\subsection{Feature Extraction and Dissimilarity Measurement}
\label{subsec_tok_vec}

% Categorizing bot loaders requires a quantification method to measure the similarity of collected request logs.

\begin{figure}
  \vspace{1px}
  \centering
  \includegraphics[width=0.7\linewidth]{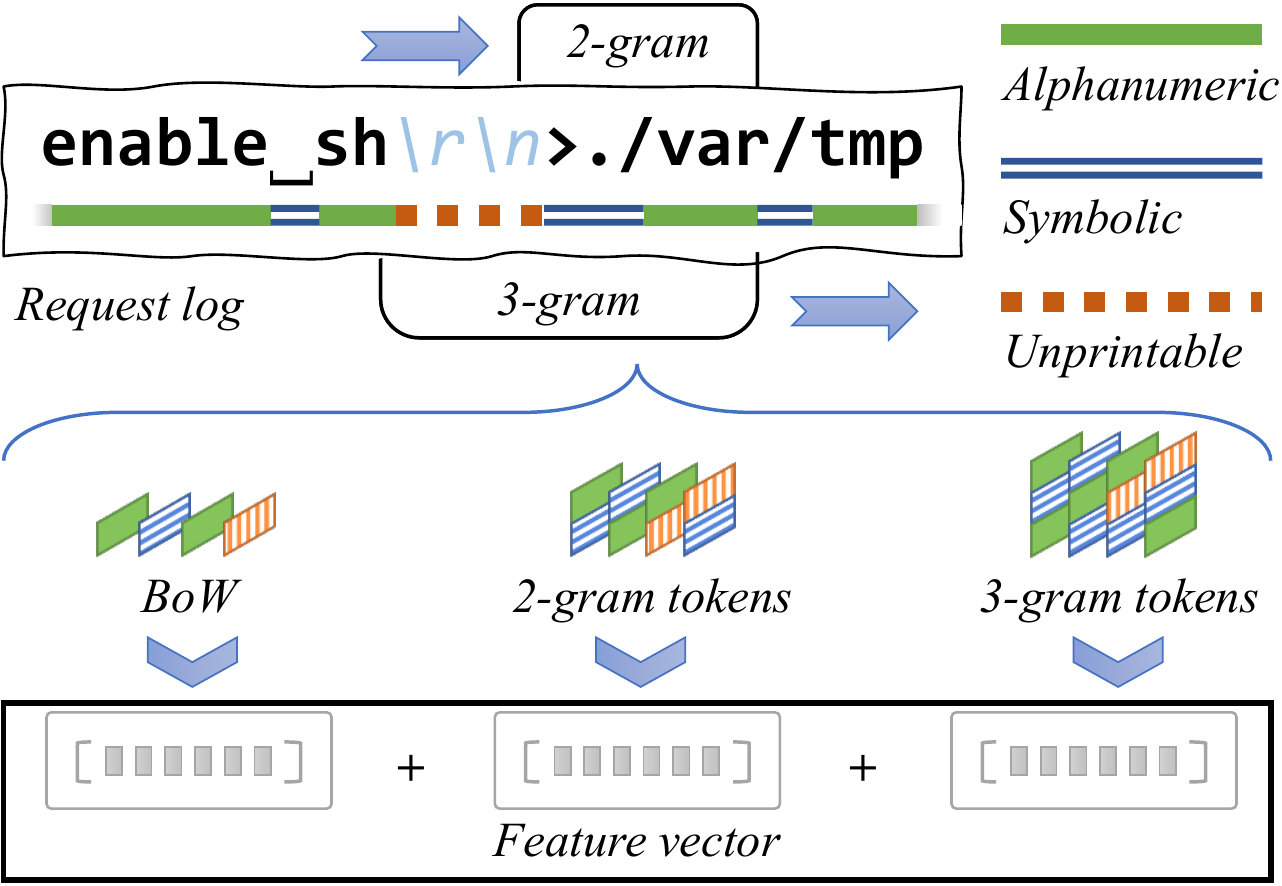}
  \caption{Byte-based tokenization and n-gram vectorization of request logs.}
  \label{fig_tokenize}
  %\vspace{-5pt}
\end{figure}

In this step, We adopt classical methods from Natural Language Processing (NLP) to embed request logs into a feature space, then quantify the dissimilarity between any two of them.
Fig. \ref{fig_tokenize} illustrates the process of feature extraction.

Similar to our work, PRISMA\cite{kruegerLearningStatefulModels2012} used bytewise 3-gram vectors to represent binary messages and token vectors to represent text messages.
However, because a single telnet message may carry both binary and text contents, we need a better embedding method to adapt to the complexity of telnet protocol.
We empirically categorize payload bytes into three types: alphanumeric, symbolic (plus punctuation and spaces), and unprintable.
As we assume that the type of each byte and its collocation imply semantic information, we split the request log at positions where two contiguous bytes are different types.
We consider these tokens as minimum semantic units of a request log and build a Bag of Word (BoW) vector to represent its basic semantics in the feature space.
This method generates a token that consists of only one type of byte and enables the extraction of information from all bytes.

Besides BoW vectors, we also use n-gram vectors to highlight the replacement of variable tokens in different loaders.
Assigning a high value to n may result in computational overhead, so we choose 2-gram and 3-gram of tokens to capture the variability while limiting the scale of feature vectors.
We join these three vectors to generate a feature vector for every request log.

In order to measure the semantic distance between request logs, the dissimilarity metric should reflect the existence, repetition, and collocation of tokens and n-gram features based on the proposed feature vector.
We choose the Euclidean distance in our experiment, because the commonly used cosine distance may not reflect the repetition of tokens.

\subsection{Agglomerative Clustering}

To cluster similar loaders yet demonstrate their inter-cluster similarities, we use agglomerative clustering to build hierarchy clusterings bottom-top, based on dissimilarity metrics.
The clustering process starts from single-element clusters corresponding to every request log.
In each iteration, the algorithm searches for two clusters having the minimum inter-cluster distance based on the distance metric (also known as the dissimilarity metric in this work) and a linkage criterion.
The algorithm hierarchically merges two clusters in each iteration until only one is left.

Here, we describe the hierarchical clusters as an inverted binary tree, on which the leaves at the bottom correspond to request logs, while a trunk node refers to an agglomerative cluster of attached leaves.
The height of a cluster node refers to the inter-cluster distance of its two sub-clusters, which also indicates the discrepancy of contained elements.
We denote agglomerative clusters as \(C\in\mathbb{N}\), where \(\mathbb{N}\) denotes the full set of them.
Every cluster \(C\) can be further split into two sub-clusters \(\{C_{1},C_{2}\}\) or merged into a super-cluster \(C^S\).
Cutting the tree at a given height \(\mathcal{T}\) will produce a preliminary partitioning \(\mathbb{P}\subset\mathbb{N}\) at a selected precision.
To make generated clusters cohesive yet discrete from each other, we determine the value of \(\mathcal{T}\) based on the shape of the tree as discussed later.
In this work, we use the \code{ward} \cite{wardHierarchicalGroupingOptimize1963} function offered by Scikit-learn, as the linkage criterion minimizes the variance of the merged clusters.

\subsection{Pattern Extraction}
\label{subsec_template}

\begin{figure}
  \vspace{1px}
  \centering
  \includegraphics[width=0.8\linewidth]{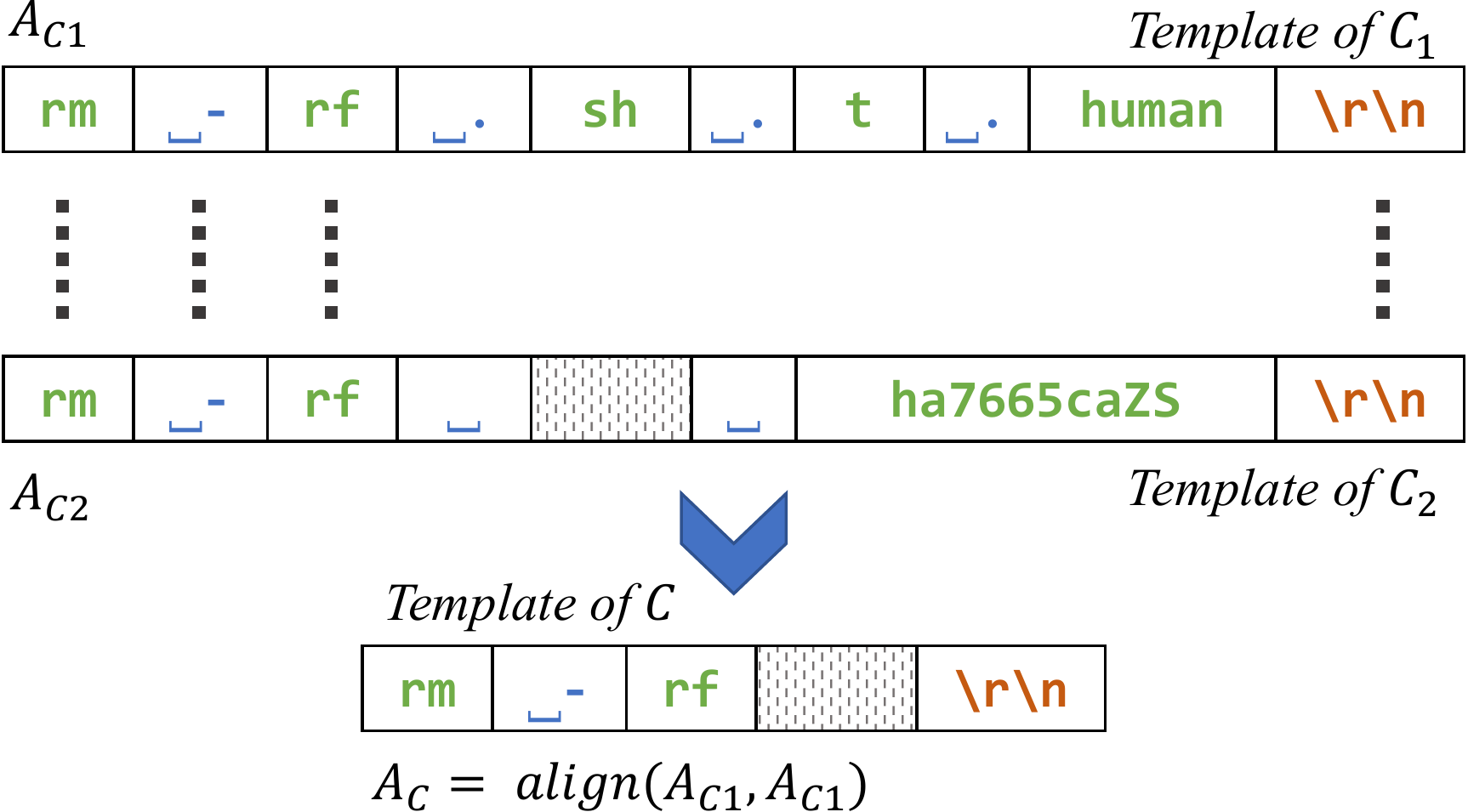}
  \caption{Smith-Waterman algorithm on the agglomerative tree.}
  \label{fig_smith_waterman}
  %\vspace{-5pt}
\end{figure}

We apply the Smith-Waterman algorithm\cite{smithIdentificationCommonMolecular1981} from leaves to the root node to get ``templates'' for every cluster on the agglomerative binary tree and identify shared patterns of sibling loaders or clusters out of their request logs.

We use the tokenized sequence in the section \ref{subsec_tok_vec} to align two request logs.
As depicted in Fig. \ref{fig_smith_waterman}, the \(align(A_{C1}, A_{C2})\) operation leverages the Smith-Waterman algorithm to scan two token sequences from head to tail.
This operation aligns identical tokens at the same position and adds placeholders (shadow cells in Fig.~\ref{fig_smith_waterman}) to replace the mismatched ones, allowing identical tokens to align.
We finally get a ``template'' of two clusters indicating the shared pattern of tokens.
For any cluster \(C\in\mathbb{N}\), the corresponding template \(A_{C}\) is generated recursively based on the templates of its sub-clusters \(A_{C1}\) and \(A_{C2}\).
Every node on the agglomerative tree will get a ``template'' representing the common pattern of its elements.

\subsection{Clustering Refinement}

As the unique \(\mathcal{T}\) value may not fit all branches on the tree, the preliminary partitioning \(\mathbb{P}\) is far from being taken as the final class definition.
Starting from nodes in \(\mathbb{P}\), we examine corresponding templates to calibrate the family definition by evaluating if a cluster should be kept, merged, or further split.
Here, we empirically configure some criteria for accepting or denying a merged cluster:

\label{manual_criteria}
\begin{itemize}
  \item While identical commands are critical in evaluating the similarity,  their arguments and arrangement are also important factors that we should concern about.
  \item For complex statements, the syntax structure is more important than its component commands to evaluate the similarity of two templates.
  \item We ignore the variation of self-identification tokens unless it appears in different commands or arguments.
\end{itemize}

% Finally, for every host we observed, we collect all related samples and their class labels, then pick the label with the maximum count as the label of this loader host.

\section{Data Analysis}
\label{sec_eval}
% !TeX root = ..\main.tex
In this section, we discuss the functions and behavior of active loaders based on the aforementioned methods and make a conclusion about their homologies.

% \begin{figure}[tpb]
%   \centering
%   \includegraphics[width=0.9\linewidth]{figures/geo_dist.pdf}
%   \caption{Geographical distribution of loader hosts.}
%   \label{fig_geo_dist_map}
% \end{figure}

\subsection{Captured Dataset}
The following analysis is based on captured request logs from November 14 to December 31 in 2021.
To reduce the scale of the dataset, we take no more than 20 request logs for each host and selected 4,855 out of over 3 million captured items.
As this work focuses on the function of active loaders instead of their deployment, we drop duplicates and got 481 valid items.
This dataset generates 895 tokens, 2,451 2-gram terms, and 4,737 3-gram terms, finally composing feature vectors of 8,083 dimensions.

% \input{tables/geoip.tex}

% According to the geographical statistics, China, Russia, India, and the United States are preferred countries for botmasters to setup loader servers.
% Over 75\% loaders are located in these 4 countries.
% While it is common sense for Mirai-based botnets to build infrastructures on public cloud servers, we can still see there are lots of servers deployed in consumer networks like ASN4837 and ASN17488.
% This means that some botnets might have a different architecture from Mirai, and loaders are also distributed across multiple networks and regions.

\subsection{Family Definition}

\begin{figure}[tpb]
  \centering
  \includegraphics[width=\linewidth]{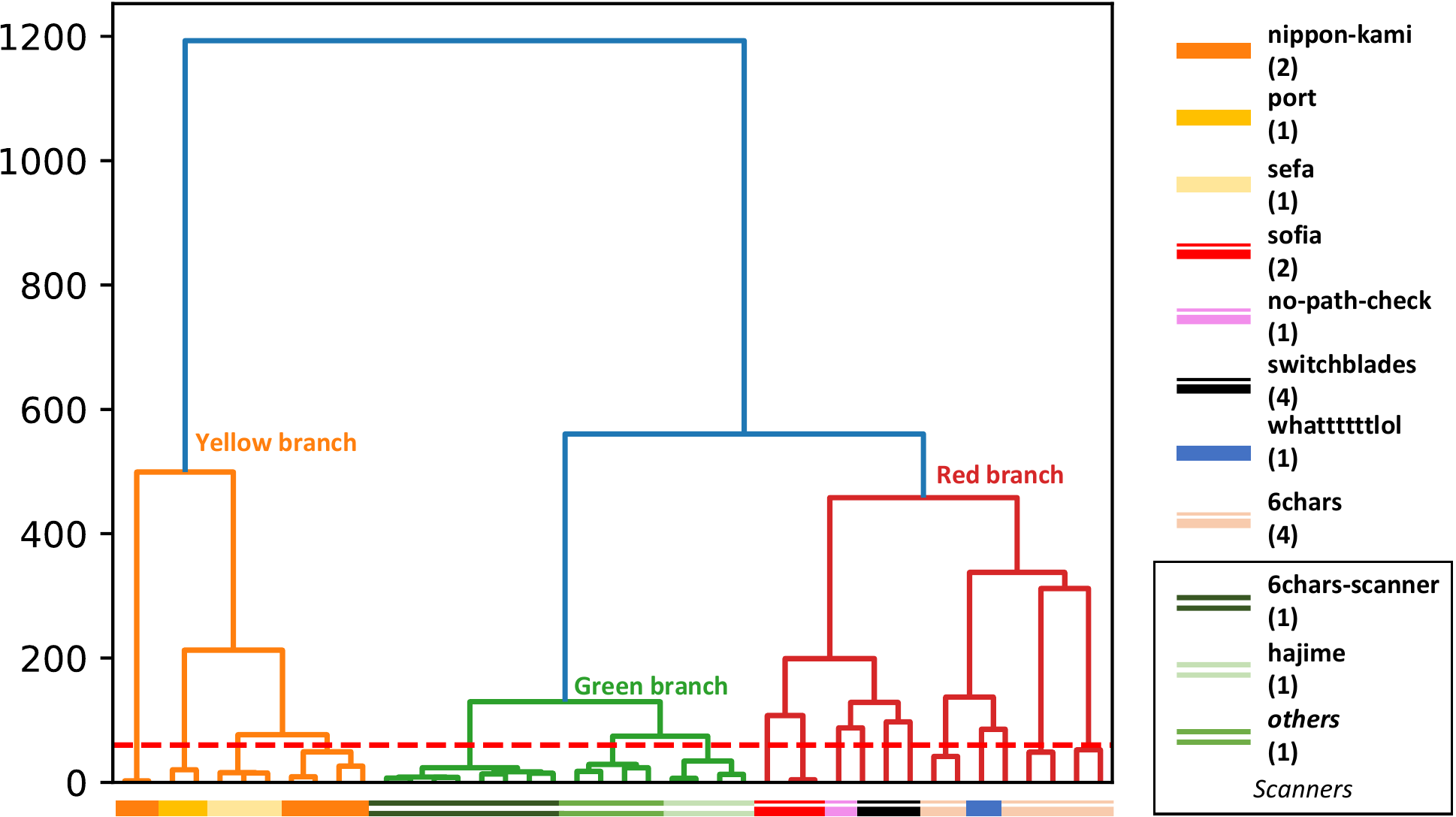}
  \caption{Clustering dendrogram of the agglomerative clustering. Only the top five layers are shown. The red dashed line denotes \begin{math}\mathcal{T}=60\end{math}. The colored bars indicate the family definition on the agglomerative tree. For every family, we note the count of member clusters/samples  in the legend on the right side.}
  \label{fig_dendrogram}
  %\vspace{-5pt}
\end{figure}

In this step, we leverage agglomerative clustering to define several families of bot loaders based on the collected dataset.

\subsubsection{Clustering overview}

The agglomerative clustering algorithm generates a tree with a height of 1193.07, whose dendrogram is depicted in \ref{fig_dendrogram}.
While the tree is relatively tall, most of the branches are at a height below 200.
Minority branches at a higher height manifest significant discrepancies in samples in the corresponding clusters.
Based on the method in Section \ref{subsec_template}, we recursively generate templates for 480 non-singleton clusters to describe their common behaviors.

\subsubsection{Threshold selection}

According to the dendrogram in Fig. \ref{fig_dendrogram}, when trying to merge two sibling clusters with a distance over 100, the intra-cluster discrepancy of the merged cluster will increase greatly compared to the original ones, which runs counter to our expectation of clustering results.
Based on this observation, we empirically set \begin{math}\mathcal{T}=60\end{math} with reasonable margin.
The \(\mathcal{T}\) value partitioned all request logs into 19 clusters.
As Griffioen's work \cite{griffioenExaminingMiraiBattle2020} only intensively investigated 14 active botnets, we regard the partitioning as reasonable to reflect the situation of active loaders.

\subsubsection{Family definition}

\begin{table}[t]
  {\centering
  \caption{Index Table of Loader Functions Discussed in Section \ref{subsec_behavior}}
  \includegraphics[width=\linewidth]{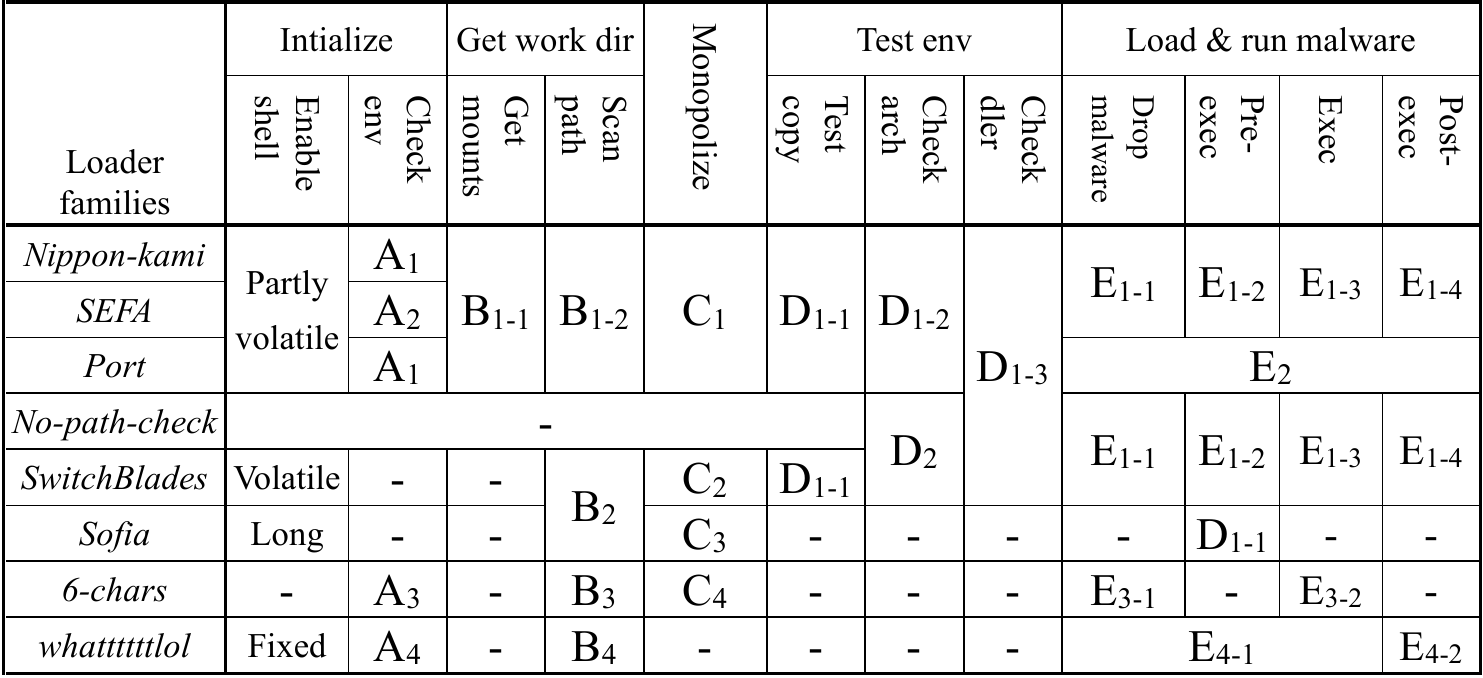}
  \label{fig_table_func}}
  * The capital letters A-E correspond to the 5 categories of loader functions listed in the upper header. Their different implementations are indicated by the first digit of their subscripts, while the second digit indicates sub-commands called for the corresponding function. 
  %\vspace{-5pt}
\end{table}

Based on the extracted templates, we evaluated these clusters based on the aforementioned criteria in section \ref{manual_criteria}, and finally identified several families out of the dataset.
We traverse their sub-clusters and pick some representative tokens as their name, which may not follow the common naming rules.
We listed these families and indexed their functions in Table \ref{fig_table_func}.

The yellow branch in Fig. \ref{fig_dendrogram} consists of 3 very similar families: \emph{Nippon-kami}, \emph{SEFA}, and \emph{Port}.
As the control group data from Anna-senpai's loader is all located on the \emph{Nippon-kami} branch, in our following analysis we treat \emph{Nippon-kami} as an alias of the original Mirai loader family.
% We also call them ``yellow families'' in the rest part of this paper.

We also identified five ``red families'' in Fig.~\ref{fig_dendrogram} that are significantly different from the aforementioned ``yellow families'':
\begin{itemize}
  \item \emph{No-path-check} removes every command prior to checking the architecture and simply uses the default working directory of the logged-in user;
  \item \emph{SwitchBlades} derives the framework of \emph{Nippon-kami}, but uses a different implementation to detect writable directory, acting similar to \emph{Sofia};
  \item \emph{Sofia} bases its intrusion toolkit on a simplified workflow, using a long initial command list and implementing a different method to detect writable directories;
  \item \emph{6-chars} generates 6 random escaped characters to check the shell environment for every session, acting differently from every other family;
  \item \emph{``whattttttlol''} does not share any pattern with other families. It runs a fixed command list and downloads multiple scripts named ``\code{whattttttlol*.sh}'' to load the bots.
\end{itemize}

\subsubsection{Clustering of scanners}
% \todo[inline]{Rephrase this paragraph. I have an intuition about what you mean, but it's oddly phrased, needs polishing.}
The samples on the green branch are very different from the other ones.
As they only conduct quick probes and do not run any downloading commands, we assume that these logs are related to scanning campaigns.
Among these scanner logs, we first identify the \emph{6-chars-scanner} family that generates 6 random escaped characters in their scanning conversation.
We also identify the \emph{hajime} family related to our control group sample generated by a Hajime bot, as well as the \emph{others} family having no particular patterns.

\subsection{Behavior Patterns}
\label{subsec_behavior}
\def\subscr#1#2{#1\textsubscript{#2}}

\lstset{
basicstyle=\scriptsize\ttfamily,
columns=flexible,
breaklines=true,
% numbers=left,
% numbersep=-10pt,
frame=single
}

\begin{figure}[t]
  % \vspace{-5px}
  \emph{\footnotesize\subscr B{1-2}}
  \vspace{-3px}
  \begin{lstlisting}
busybox echo -e '\\x6b\\x61\\x6d\\x69/proc' > /proc/.nippon; 
busybox cat /proc/.nippon; 
busybox rm /proc/.nippon
  \end{lstlisting}
  \vspace{-5px}
  \emph{\footnotesize\subscr B{2} and \subscr B{3}}
  \vspace{-3px}
  \begin{lstlisting}
>/var/tmp/.file && cd /var/tmp/
  \end{lstlisting}
  \vspace{-5px}
  \emph{\footnotesize\subscr D{1-1}}
  \vspace{-3px}
  \begin{lstlisting}
/bin/busybox cp /bin/echo sefaexecbi; >sefaexecbi; /bin/busybox chmod 777 sefaexecbi; 
  \end{lstlisting}
  \vspace{-5px}
  \emph{\footnotesize\subscr D{1-2}}
  \vspace{-3px}
  \begin{lstlisting}
/bin/busybox cat /bin/echo
  \end{lstlisting}
  \vspace{-5px}
  \emph{\footnotesize\subscr D{2}}
  \vspace{-3px}
  \begin{lstlisting}
/bin/busybox cat /bin/busybox || while read i; do echo $i; done < /bin/busybox
  \end{lstlisting}
  \vspace{-5px}
  \caption{Sample codes of the ``Get working directory'' function (\subscr B{*}) and the ``Test environment'' function (\subscr D{*}) denoted by indexes in Table \ref{fig_table_func}.}
  \label{codefig}
  %\vspace{-5pt}
\end{figure}

In this section, we interpret the Table \ref{fig_table_func} vertically to make a comprehensive comparison about their shared pattern.
We denote all functions of loaders by alphanumeric indexes.

\subsubsection{Initialize} 
At the beginning of the intrusion, the loader injects initializing commands to enable the shell interface and checks the environment.
As yellow families share the same codebase, their initialize command lists are very similar.
They run \code{ps} command to check suspicious processes in the environment (\subscr A{1}).
The SEFA loader modifies the victim's hostname to \code{SEFA\_ID:\textit{<4-digit numbers>}} (\subscr A{2}) to identify bots in the botnet.
Sofia removed all checking commands but extended the initialize command list.
While whattttttlol holds a fixed command list, it runs \code{ls /home} to scan files in the directory (\subscr A{3}).
6-chars only checks \code{wget} in this step (\subscr A{4}).

\subsubsection{Get working directory}

Most of the loaders require a writable directory to temporarily drop the executable.
Yellow families scan mounted filesystems (\subscr B{1-1}) and create some files(\subscr B{1-2}) to check their writing privileges.
%TODO 表述可能有错！
% \todo[inline]{This is unclear. B2 and B3 are used by these three malware, or is the yellow families that use B2 and B3? There is some confusion of the subject. Rephrase to make it clear who does what, and what not. In subsection "Monopolize" you did good, because it's clear who does what.}
SwitchBlades, Sofia, and 6-chars use a simplified statement (\subscr B{2} and \subscr B{3} in Fig. \ref{codefig}) to test writable directories in their own hard-coded lists.
While SwitchBlades and Sofia use returns to assemble these element statements (\subscr B{2}), the 6-chars family uses semicolons (\subscr B{3}) which makes a slight difference.
A variant of 6-chars runs this step twice, which shows a difference in the request logs.
Whattttttlol uses a simple ``\code{||}'' (or) statement to join multiple \code{cd \textit{<directory>}} commands (\subscr B{4}).
This statement changes the working directory to the first available one in the hard-coded list, regardless of its writable privilege.

\subsubsection{Monopolize} 
Most loaders will try eliminating competitors by deleting certain files stored in a built-in list.
The yellow families use \code{.sh .t .human} (\subscr C{1}), Sofia uses \code{.file .cowbot.bin retrieve cowffxxna} (\subscr C{3}), and 6-chars uses \code{.i} only (\subscr C{4}).
SwitchBlades tries to delete two files while the lists are unstable among different variants (\subscr C{2}).

\subsubsection{Test environment}

In this step, the loaders probe the CPU architecture and the available downloaders to decide how to load a bot.
Yellow families tests \code{cp} command(\subscr D{1-1}), prints \code{/bin/echo}(\subscr D{1-2}), and then test \code{wget} and \code{tftp} commands (\subscr D{1-3}) in this step.
As the CPU architecture can be obtained by parsing any executable on the device, no-path-check and Sofia prints \code{/bin/busybox} to obtain the same information (\subscr D{2}).
In case of the \code{cat} command is unavailable, they also use a shell-based \code{while read} statement to print the file.
Sample codes are displayed in Fig. \ref{codefig}.

\subsubsection{Drop \& run malware} 
In this step, loaders \code{cd} to the selected working directory and drop bot clients via a tested downloader.
If neither \code{wget} nor \code{tftp} is available, most Mirai-based families will run a fallback command that loads the whole file with \code{echo} command and launches a stdout redirect statement (\subscr E{1-*}).
6-chars leverages an ``\code{||}'' (or) statement to call multiple commands sequentially (\subscr E{3-*}) until a command succeeds.
Whattttttlol calls multiple commands sequentially to download and run 4 scripts, and then it deletes them all after the execution to clean the trace (\subscr E{4-*}).
In this step, Port and Sofia do not seem to download any executable.
Instead, Port calls \code{openssl} for an unknown reason (\subscr E{2}), while Sofia only checks writable privilege in the current directory (\subscr D{1-1}).

\section{Discussion}
\label{sec_discussion}
% !TeX root = ..\main.tex

\begin{figure*}[htbp]
    \centering
    \includegraphics[width=0.9\textwidth]{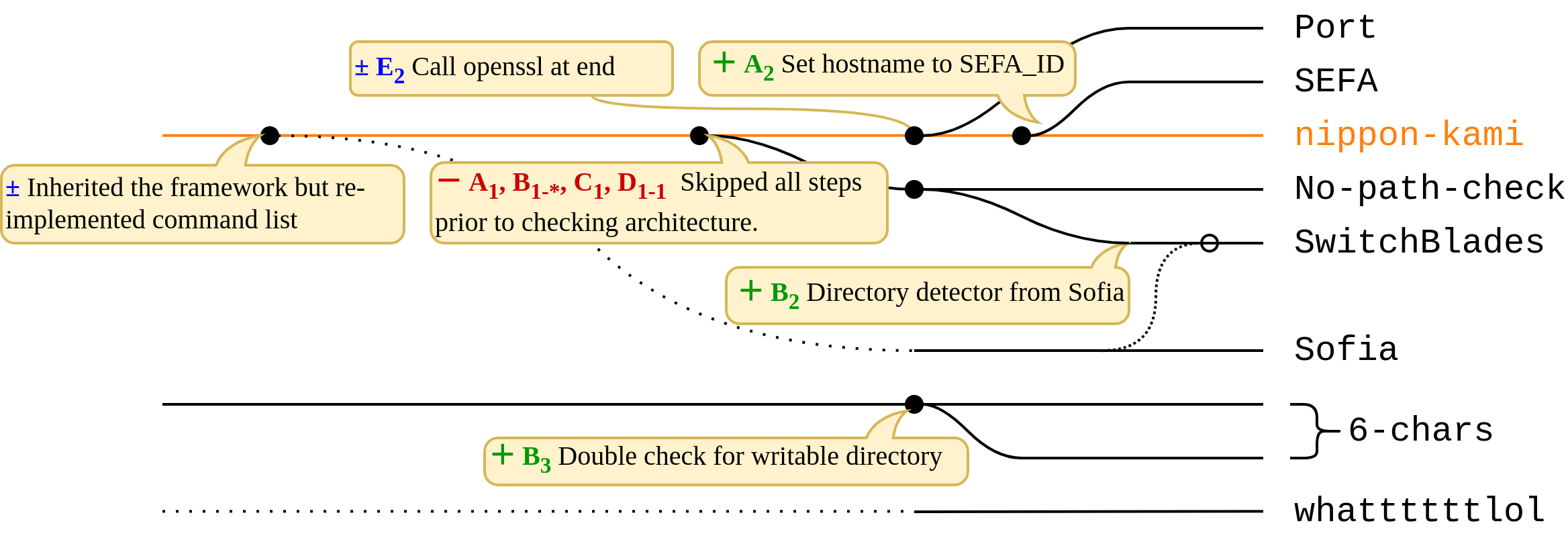}
    \caption{Dendrogram of loader families. Horizon lines depict identified families, while the yellow one highlights the \emph{nippon-kami} released by Anna-Senpai (the author of Mirai). In the description of a connecting line between two families, green ``+'' indicates adding functions, red ``-'' indicates removing functions, and blue ``\textpm'' indicates modifications. All the functions are denoted by indexes in Table \ref{fig_table_func}.}
    \label{fig_dendro_manual}
    \vspace{-10pt}
\end{figure*}

Based on the agglomerative tree and the discussion about behavior patterns, we draw a dendrogram (Fig. \ref{fig_dendro_manual}) to demonstrate the variation of intrusion functions.

\subsection{Variation of Loaders}
Although we treat \emph{nippon-kami} family as the direct descendant of the original Mirai loader, we found that other families inherit its intrusion workflow, but modify some components to adapt to different environments and situations.
The directory detector is frequently modified or rebuilt to fit heterogeneous filesystem structures on victim devices, while the cleaning commands used for monopolizing the infected device also vary according to the malware family.
Some families made significant changes to the original code base to simplify the workflow (\emph{No-path-check} and \emph{SwitchBlades}) or rebuild the toolkit (\emph{Sofia}).
Some independent families also implemented their toolkits to load malware.

While conventional taxonomy research overlooked the variation and evolution of bot loaders, this experiment reveals that Mirai original ideas and codebase are still contributing to new spawning variants. Our server-side perspective highlights how the infection mechanisms of these bots operate through telnet and how they are suitable to run on different environments.

\subsection{Comparison with Other Studies}
According to Cozzi\cite{cozziTangledGenealogyIoT2020} and Wang~\cite{wangEvolutionaryStudyIoT2021}, the evolution of bot clients focused on updating their scanners, attackers, persistence techniques, and anti-detection techniques.
Our work demonstrated the distinct motivation of loaders' and bots' evolution. 
% \todo[]{This phrase "Our work has revealed the distinct reason and preference for loaders' evolution. " is cryptic. Rephrase.}
Compared to Torabi's work~\cite{torabiStringsBasedSimilarityAnalysis2021} and Tabari's work~\cite{tabariWhatAreAttackers2021}, our work further quantified the similarity of families and identified the lineage of loaders beyond simple comparisons of string patterns, which contributes to understanding the evolution of botnet malware from new perspectives.
% \todo[]{Also this is cryptic, rephrase.}

As noted by Wang~\cite{wangEvolutionaryStudyIoT2021} and Dang~\cite{dangUnderstandingFilelessAttacks2019}, a growing number of botnets are exploiting victims by means of fileless attacks.
We noticed the \emph{Port} family replaced \emph{Nippon-kami}'s loading tool with a fileless attack command, which broke Mirai convention behaviour of infecting victims by downloading executables.
Our lineage study sheds some light on the provenance of fileless attack toolkits and helps to understand how botmasters develop new attack vectors starting from the original Mirai codebase. In turn, this knowledge can contribute to improve the efficiency of defense strategies against new variants.

\section{Conclusion}
\label{sec_conclusion}
% !TeX root = ..\main.tex
% Besides the thorough investigation of bot clients, loaders are not systematically investigated due to the deficiency of existing methods.
In this paper, we analysed telnet request logs captured with ad-hoc honeypots, and we investigated functions and similarities of various infection loaders. Our data allowed us to define 8 different families and draw a dendrogram of their lineage and evolution, demonstrating the importance of understanding loaders' evolution and variation.
The experiment highlighted the evolution of IoT botnets on the server side, providing a server-side view of botnets evolution and a novel behavior-based taxonomy of bot loaders.
% \todo[]{Conclusion needs a little more info. You can take some of the information written in the abstract and replicate it here. The conclusion should be written with the spirit of "In abstract we promised these things, here I recap and we show that we kept our promise".}

\balance
\bibliographystyle{IEEEtran}
\bibliography{IEEEabrv, lib}

\end{document}